\documentclass[a4paper]{cas-sc}
\usepackage[numbers]{natbib}
\usepackage{multicol}

\def\tsc#1{\csdef{#1}{\textsc{\lowercase{#1}}\xspace}}
\tsc{WGM}
\tsc{QE}
\begin{document}

\let\WriteBookmarks\relax
\def\floatpagepagefraction{1}
\def\textpagefraction{.001}

\newcommand{\WSe}{$\text{WSe}_{2}$\xspace}
\newcommand{\MoSe}{$\text{MoSe}_{2}$\xspace}
% Short title
\shorttitle{Optical response of \WSe-based vertical tunneling junction}    

% Short author
\shortauthors{K. Walczyk et al.}  

% Main title of the paper
\title[mode = title]{Optical response of WSe$_2$-based vertical tunneling junction}  

\author[1]{K. Walczyk}%[]
\credit{Investigation, Data curation, Writing - Original draft preparation}

\cormark[1]
\cortext[cor1]{Corresponding authors}
\ead{kp.walczyk@student.uw.edu.pl}

\author[1]{G. Krasucki}%[]
\credit{Investigation, Data curation}

\author[1]{K. Olkowska-Pucko}%[]
\credit{Investigation, Data curation}

\author[3]{Z. Chen}%[]
\credit{Sample fabrication}

\author[5]{T. Taniguchi}%[]
\credit{Provide the hBN crystals}

\author[4]{K. Watanabe}%[]
\credit{Provide the hBN crystals}

\author[1]{A. Babiński}%[]
\credit{Conceptualization}

\author[2,3]{M. Koperski}%[]
\credit{Conceptualization}

\author[1]{M. R. Molas}%[]
\credit{Conceptualization, Funding Acquisition, Project administration, Writing - Original draft preparation}

\cormark[1]
\ead{maciej.molas@fuw.edu.pl}

\author[1]{N. Zawadzka}%[]
\credit{Conceptualization, Project administration, Writing - Original draft preparation}

\cormark[1]
\ead{natalia.zawadzka@fuw.edu.pl}

% Address/affiliation
\affiliation[1]{organization={Institute of Experimental Physics, Faculty of Physics, University of Warsaw},
           % addressline={Ludwika Pasteura 5}, 
            city={Warsaw},
%          citysep={}, % Uncomment if no comma needed between city and postcode
            postcode={02-093}, 
           % state={},
            country={Poland}}

\affiliation[2]{organization={Dep. of Materials Science and Engineering, National University of Singapore},
           % addressline={}, 
            city={Singapore},
%          citysep={}, % Uncomment if no comma needed between city and postcode
            postcode={117575}, 
           % state={},
            country={Singapore}}            

\affiliation[3]{organization={Inst. for Functional Intelligent Materials, National University of Singapore},
           % addressline={}, 
            city={Singapore},
%          citysep={}, % Uncomment if no comma needed between city and postcode
            postcode={117575}, 
          %  state={},
            country={Singapore}} 

\affiliation[4]{organization={Research Center for Electronic and Optical Materials, National Institute for Materials Science},
           % addressline={1-1 Namiki}, 
            city={Tsukuba},
%          citysep={}, % Uncomment if no comma needed between city and postcode
            postcode={305-0044}, 
          %  state={},
            country={Japan}}

\affiliation[5]{organization={Research Center for Materials Nanoarchitectonics, National Institute for Materials Science},
         %   addressline={1-1 Namiki}, 
            city={Tsukuba},
%          citysep={}, % Uncomment if no comma needed between city and postcode
            postcode={305-0044}, 
          %  state={},
            country={Japan}}

\begin{abstract}
Layered materials have attracted significant interest because of their unique properties.
Van der Waals heterostructures based on transition-metal dichalcogenides have been extensively studied because of potential optoelectronic applications. 
We investigate the optical response of a light-emitting tunneling structure based on a WSe\textsubscript{2} monolayer as an active emission material using the photoluminescence (PL) and electroluminescence (EL) experiments performed at low temperature of 5~K.
We found that the application of the bias voltage allows us to change both a sign and a value of free carriers concentrations.
Consequently, we address the several excitonic complexes emerging in PL spectra under applied bias voltage. 
The EL signal was also detected and ascribed to the emission in a high-carrier-concentration regime. 
The results show that the excitation mechanisms in the PL and EL are different, resulting in various emissions in both types of experimental techniques.

\end{abstract}


\maketitle

% Main text

\section{Introduction}

Layered van der Waals heterostructures based on transition metal dichalcogenide (TMD) monolayers (MLs) attract great attention from the community due to their potential optoelectronic applications~\cite{Wang2020, Malic2018, Goki2018, Lien2018, Babar2023}. 
MLs of molybdenum- and tungsten-composed TMDs, $i.e.$ MoS$_2$, MoSe$_2$, MoTe$_2$, WS$_2$, and WSe$_2$, are direct band-gap semiconductors, which can be divided into two subgroups: bright and darkish~\cite{Koperski2017, Koperski2019, Molas2019, Molas2017, Jadczak_2017}. 
In bright MLs, $i.e.$ MoSe$_2$ and MoTe$_2$, the energetically lowest state is optically active (bright exciton), and their low-temperature emission spectra are formed by two resonances ascribed to bright neutral and charged excitons~\cite{ Koperski2017, Molas2019, Jadczak_2017, Lu2020}.
In contrast, the energetically lowest state in darkish MLs, $i.e.$ WS$_2$ and WSe$_2$, is optically inactive (dark exciton), and thus the low-temperature emission spectra of these MLs are characterised by a variety and multiplicity of excitonic complexes, $e.g.$ neutral and charged excitons and biexcitons, dark excitons and trions, and phonon replicas ~~\cite{Arora2015, Arora2020, Zinkiewicz2020nanoscale, Zinkiewicz2022, Zinkiewicz2021nanolett, Kapuściński2021, Molas2017nano, Jadczak_ACS}.  
The optical responses of TMD MLs were extensively investigated using a photoluminescence (PL) technique~\cite{Koperski2017, Koperski2019, Molas2019, Molas2017, Jadczak_2017}, while the electroluminescence (EL) technique is less widely used in that field. The difference in the use of both types of experiments comes from the methods of carrier excitation. 
Although the PL emission emerges after excitation of the material using laser light, the EL signal appears as a result of electrical injection of carriers. 
Consequently, the structures used in the EL investigations are tunneling structures with a series of different layers working as active material ($e.g.$ a monolayer), tunneling barrier (thin layers of hexagonal BN (hBN)), and electrical electrodes (graphene). 
Characterization of EL signals enables us to study multiple materials, their heterostructures, and phenomena such as single-photon emitters and tunable emission in vertical tunneling geometry, as has been shown in recent studies~\cite{Howarth2024, Grzeszczyk2024, Zultak2020}. 
However, because EL experiments lead to a much larger population of excited carriers compared to PL, comparative studies are necessary. 

\begin{figure}
		\centering
		\includegraphics[width=0.48\textwidth]{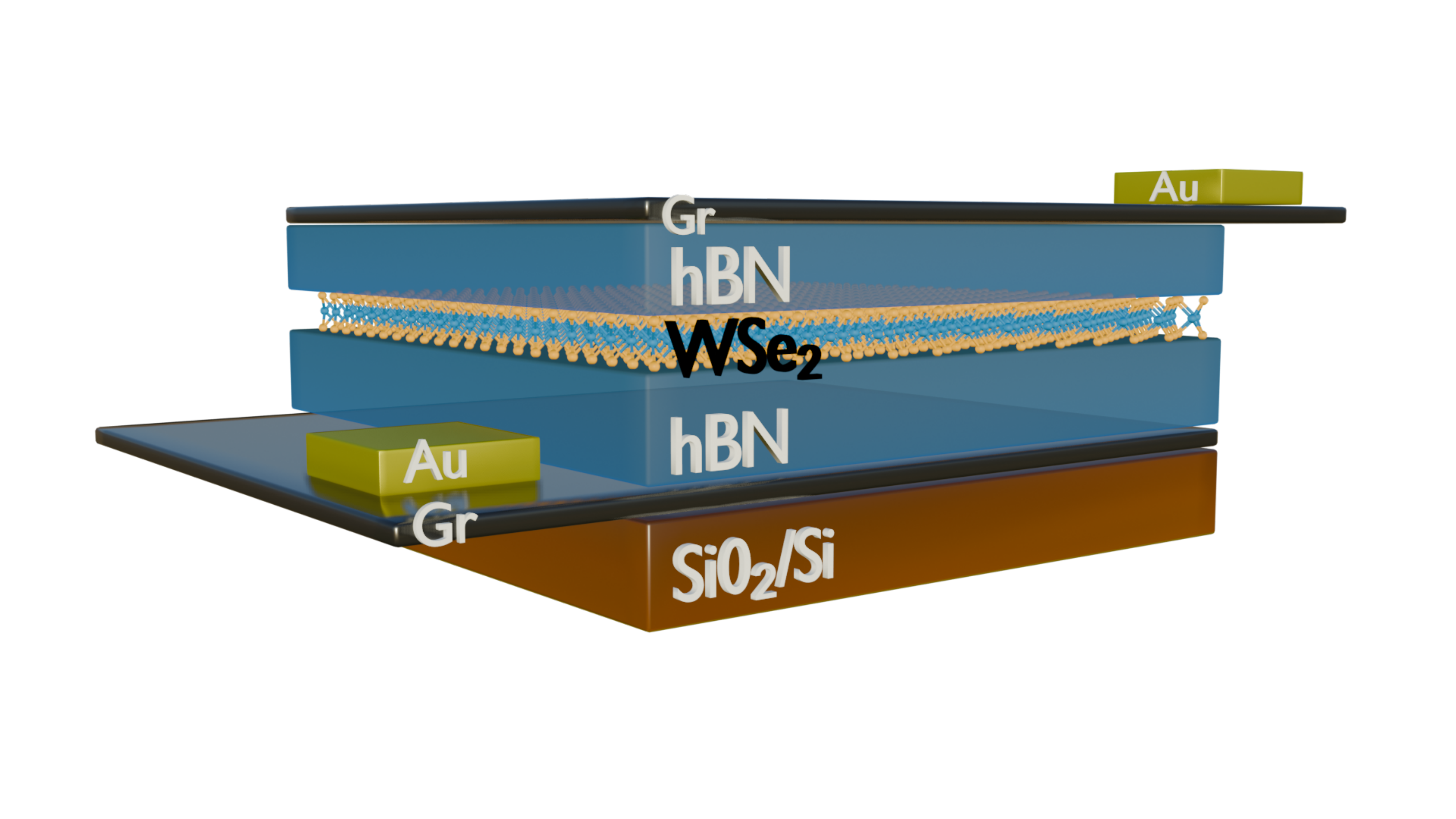}
		\caption{Schematic drawing of the investigated heterostructure with the WSe$_{2}$ ML encapsulated in the hBN flakes and embedded between the graphene (Gr) electrodes.}
		\label{fig1}
\end{figure}
		
In this work, the optical response of a light-emitting tunneling structure (see Figure~\ref{fig1}) based on a WSe\textsubscript{2} ML as an active emission material was investigated with the aid of PL and EL at a low temperature of 5~K.
A series of well-resolved emission lines seen in the low-temperature PL spectra under applied bias voltage was ascribed to different excitonic complexes, which confirms the high quality of  the examined ML.
The application of the external bias voltage allows us to significantly change the concentration of the free carriers, which was accompanied by the observation of peaks in the PL spectra related to the neutral, positively, and negatively charged complexes.
The application of bias voltages at the level of about $\pm$4~V results in emergence of the EL signal.
Because of the asymmetry of the studied structure (different thicknesses of hBN barriers), the EL spectra differ significantly when the sign of bias voltage is changed. 
Consequently, the EL signal was investigated in a high carrier concentration regime.
The heating of the sample due to the passing tunneling current is not negligible and results in a significant broadening of the resonances dominated by charged excitons.

\begin{figure}
	\centering
	\includegraphics[width=0.90\textwidth]{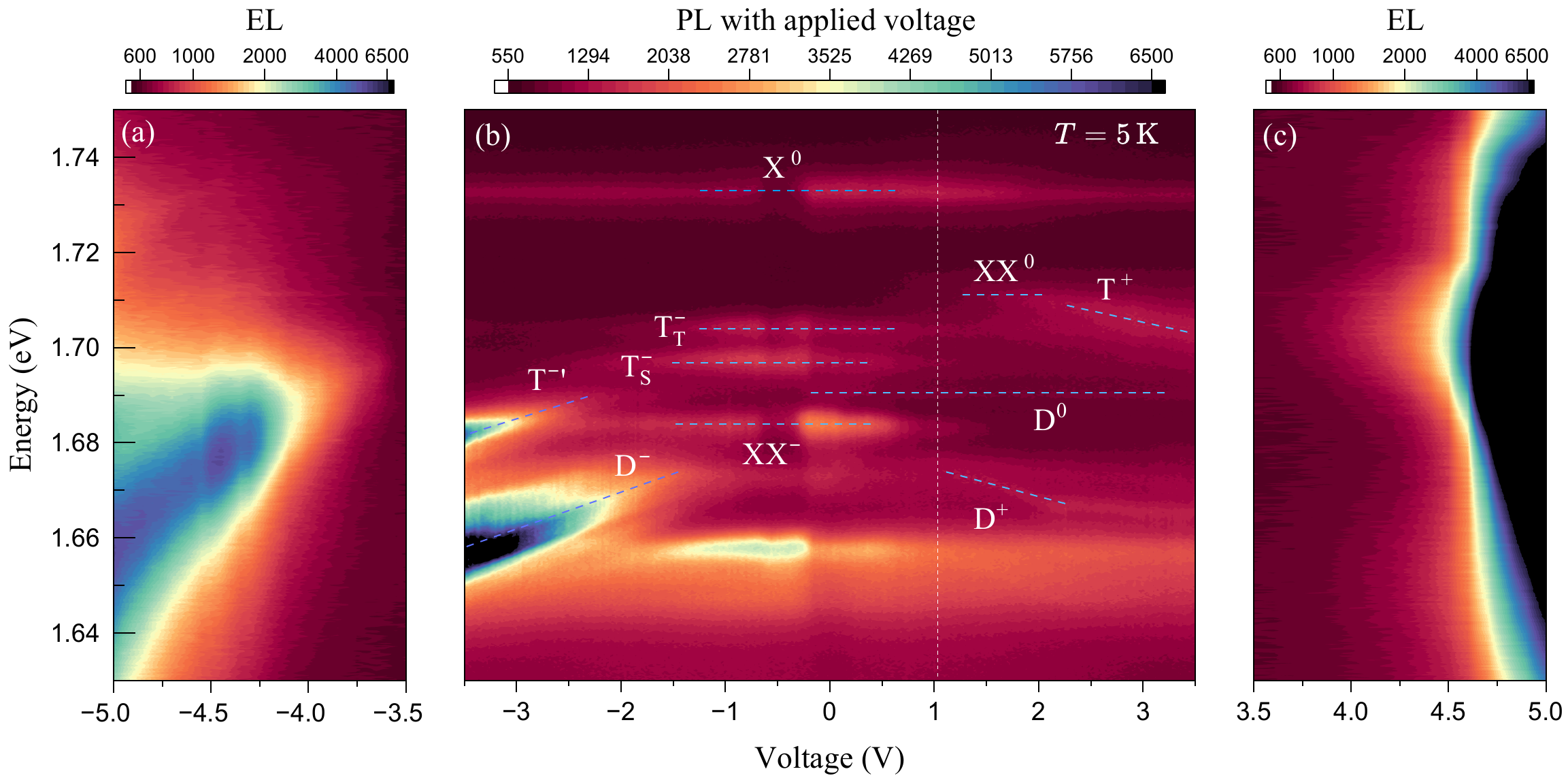}
	\caption{Voltage-dependent  (b) PL and (a, c) EL maps at 5 K. PL measurements were performed with  an excitation energy of 2.41 eV and a laser power of $\sim$ 25 $\mu$W. 
		The colour scale represents the PL and EL intensity. 
		Different identified excitonic complexes (X$^0$, T$^+$, T$^{-}_\text{T}$, T$^{-}_\text{S}$, D$^0$, D$^+$, D$^-$, XX$^0$, XX$^-$, T$^{-'} $) were marked with light blue dashed lines.
		The vertical white dashed line represents the charge neutrality point. }
	\label{fig2}
\end{figure}

\section{Methods}
\subsection{Sample}
The examined light-emitting device (LED) was fabricated using mechanical exfoliation and dry transfer methods and placed on a silicon substrate covered by a thin SiO$_2$ layer.
The heterostructure was composed of a ML of \WSe encapsulated between thin hexagonal boron nitrade (hBN) barriers with top and bottom transparent graphene electrodes, see Figure \ref{fig1}. 
As extracted by atomic force microscopy (AFM), the hBN barriers have different thicknesses equal to 7 and 3 layers in the bottom and top barrier, respectively.
After transfer, the gold edge contacts to the two graphene electrodes were attached with the aid of the ion-beam lithography technique. 
	
\subsection{PL and EL measurements}
The PL experiments were performed using $\lambda = 514.5$ nm (2.41 eV) continuous-wave (CW) laser diode. 
The studied sample was placed on a cold finger in a continuous-flow cryostat mounted on x–y motorised positioners.
The excitation light was focused by means of a 50x long-working-distance objective with a 0.55 numerical aperture producing a spot of about 1 $\mu$m diameter. 
The signal was collected via the same microscope objectives, sent through a 0.75 m monochromator, and then detected using a liquid nitrogen-cooled charge-coupled device (CCD) camera. 
For EL measurements, a sourcemeter unit (Keithley 2450) was used. 
The voltage was applied to the bottom electrode, while the top contact was grounded. 
The EL emission was collected using the same objective as that for the PL.

\section{Results and discussion}
The characteristics of the light emission of the investigated \WSe LED at 5 K is shown in Figure \ref{fig2}. 
The low-temperature PL response as a function of the applied bias voltage is demonstrated in panel (b) of the Figure, while Figures \ref{fig2}(a) and (c) present the corresponding EL signal.  
As can be seen in Figure \ref{fig2}(b), the bias-dependent PL spectra are composed of several narrow emission lines, which confirm the high quality of the studied WSe$_2$ ML \cite{Zinkiewicz2022, Molas2019, Koperski2017}.
	
\begin{figure}
		\centering
		\includegraphics[width=0.45\textwidth]{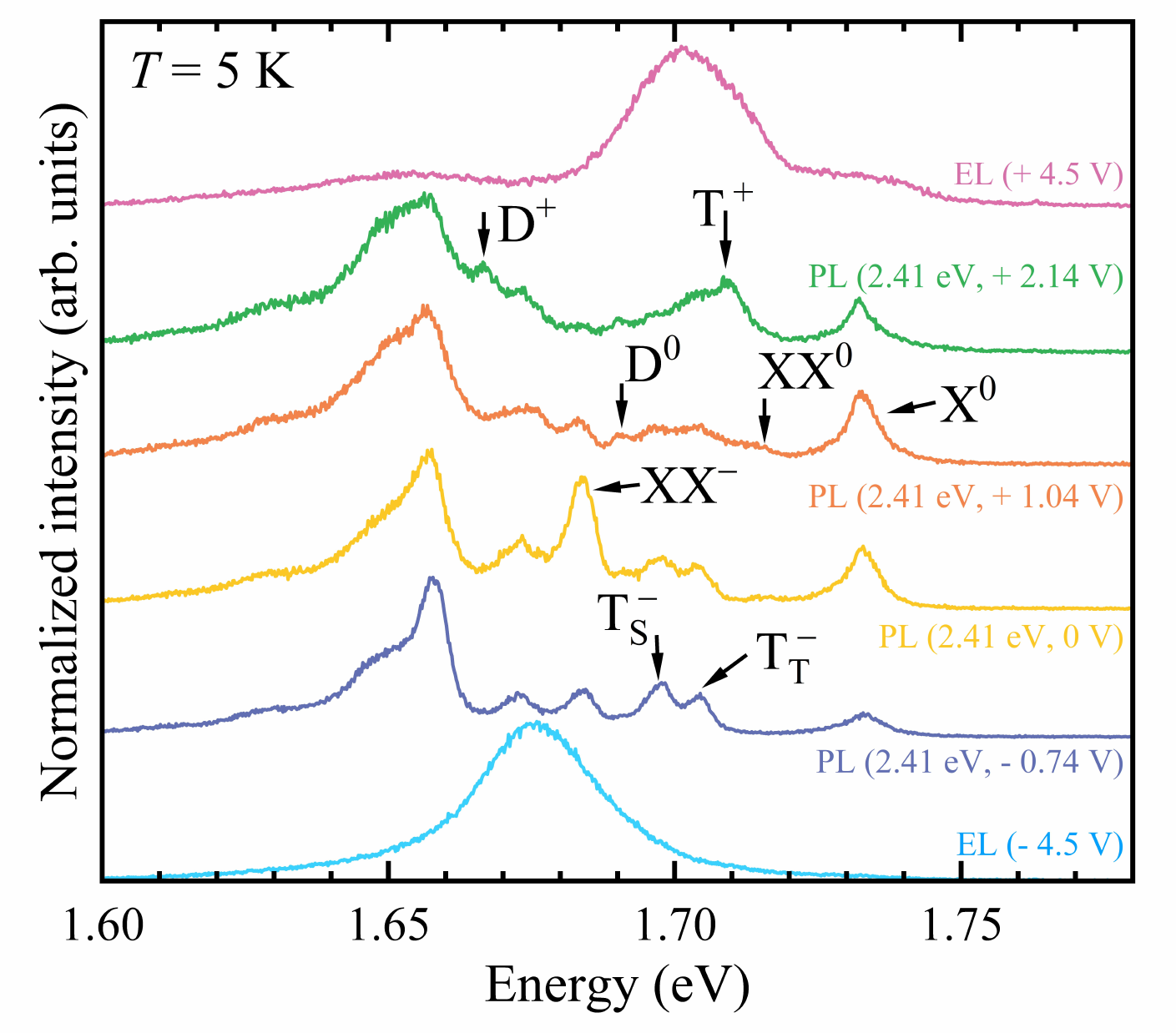}
		\caption{EL and PL spectra of the investigated sample at different voltage values. The spectra are normalized to the maximum intensity and shifted vertically for clarity with some excitonic complexes marked. }
		\label{fig3} 
\end{figure}
	
First, lets focus on the identification of the apparent emission lines.
The PL spectra display several emission lines with a characteristic pattern similar to that previously reported in several works on WSe$_2$ MLs embedded between hBN flakes~\cite{ Arora2020, Zinkiewicz2022, Li2019replica, Courtade2017, Li2018, Chen2018, Barbone2018, Liu2019, Li2019momentum, LiuValley, Liu2020, He2020, Robert2021, Wang2017}. 
Consequently, the assignment of the apparent peaks was performed according to the energy positions of excitonic complexes reported in the literature~\cite{ Arora2020, Zinkiewicz2022, Li2019replica, Courtade2017, Li2018, Chen2018, Barbone2018, Liu2019, Li2019momentum, LiuValley, Liu2020, He2020, Robert2021, Wang2017}.  
The PL spectrum measured at zero bias (see yellow curve in Figure~\ref{fig3}) is dominated by a negative biexciton (XX$^-$) with its intensity two times bigger than that of a neutral bright exciton (X$^0$).
There are also two bright negatively charged excitons (negative trions) in the spectrum, which can be ascribed to a spin-singlet (T$^{-}_\text{S}$) and a spin-triplet (T$^{-}_\text{T}$) negative trions. 
This indicates that the investigated \WSe ML at zero bias is unintentionally $n$-type doped, $i.e.$ there is a non-negligible concentration of free electrons. 
The dark intravalley spin-forbidden exciton (D$^0$) can also be observed \cite{Liu2020}.
For the whole range of applied bias voltages, there is a broad emission signal below 1.66 eV, which attribute to localised excitons and phonon replicas \cite{Zinkiewicz2022, Li2019replica}.

When the applied voltage is decreased, we increase the number of free electrons. 
Consequently, the probability to observe emission lines related to negatively charged excitonic states grows.
Two new emission lines significantly gain their intensities, which is accompanied by a quenching of the intensity of XX$^-$. 
The first one, we assign to the dark intravalley negative trion (D$^-$), the intensity of which dominates the measured PL at the smallest values of the applied voltage.
The latter peak, labelled T$^{-'}$, has been reported in several recent studies~\cite{Goryca2020, Tuan2017}, but its exact origin is still under debate.
The T$^{-'}$ line appears only in the regime of large electron densities, does not emerge in a hole-doped system, and is not observed in \MoSe ML \cite{Goryca2020, Tuan2017}.

\begin{figure}
	\centering
	\includegraphics[width=0.6\linewidth]{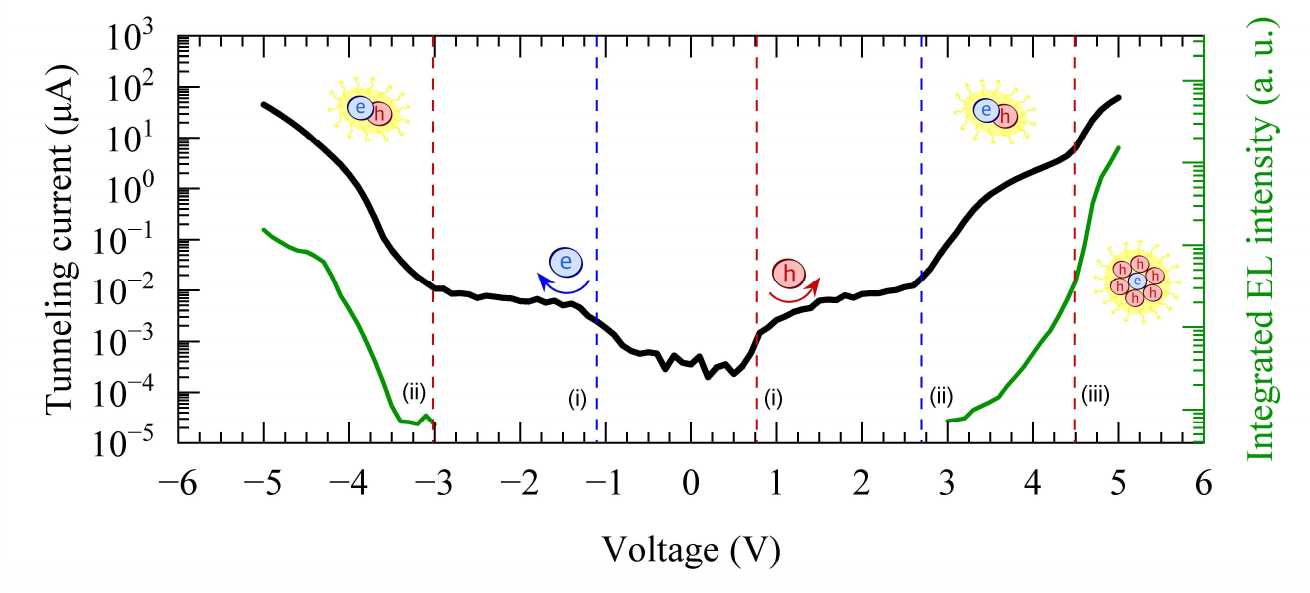}
	\caption{(black curve) The tunneling current-voltage (IV) curve and (green curve) the integrated EL intensity of our device. Both restuls are presented on a logarithmic scale. The IV curve represents data obtained under a laser excitation of 2.41~eV, $i.e.$ during measurements of the PL spectra.}
	\label{fig4}
\end{figure}

In opposite case, with an increase in positive bias voltage, the number of free electrons is substantially decreased.
This results in the observation of the charge neutrality point of our device at the level of 1.04~V, see vertical dashed line in Figure \ref{fig2}(b) and orange curve in Figure~\ref{fig3}. 
In the vicinity of this point, the emission line associated with the neutral biexciton (XX$^0$) can be seen.
The consecutive increase in bias voltage leads to a change in the doping sign to the $p$-type, which is accompanied by the emergence of free holes.
We recognise two new excitonic complexes: a bright positive trion (T$^{+}$) and a dark intravalley positive trion D$^+$, see green curve in Figure~\ref{fig3}. 
The energy differences of excitonic complexes observed in negative and positive bias voltages, $e.g.$ T$^{-}_\textrm{T/S}$ versus T$^{+}$, are due to their different binding energies, which originates from the electronic structures in both conduction and valence bands and to the short-range Coulomb exchange interaction between the charge carriers \cite{Courtade2017}. 
	
For biases |$V_{b}$|>3.5 V, the EL signal is measurable, see Figure~\ref{fig2}(a) and (c). 
As can be appreciated in Figure~\ref{fig3}, the shapes of the EL spectra at $\pm$4.5~V looks very similar (compare the dark pink and light blue curves in the figure), $i.e.$ are formed by broad emission bands centred at about 1.702~eV/1.675~eV.
The measured EL spectra are analogous to those reported in Ref.~\cite{Withers2015nano}.
Due to the high concentrations of free carriers in both limits of the applied bias voltages, we ascribe these emissions to the recombination of many-body complexes formed by a single particle charge, $e.g.$ an electron or a hole, and a sea of complementary free carriers, correspondingly holes or electrons.
	
It is evident from Figures \ref{fig2}(a) and (c) that the device shows asymmetric behaviour for positive and negative biases, which is reflected in the tunneling current-voltage (IV) curve measured on the investigated device presented in Figure~\ref{fig4}.
The figure shows the IV curve of our device obtained under a laser excitation of 2.41~eV, $i.e.$ during measurements of the PL spectra.
The faster growth of the EL intensity in positive bias voltages compared to negative ones (see Figures \ref{fig2}(a) and (c)) can be described in terms of the different thicknesses of the top and bottom hBN barriers, and therefore various tunneling of free carriers electrons or holes~~\cite{Novotny2023, Britnell2012, Grzeszczyk2024}. 
	
Let us focus on the detailed analysis of the IV curve shown in Figure~\ref{fig4}.
The shape of the IV curve for the positive and negative bias voltages is dramatically different.
In particular, the onset for the positive voltages is apparent at around 0.8~V and is more abrupt compared to the onset for the negative voltages observed at around -1~V. 
This discrepancy can be related to the different thicknesses of both hBN tunneling barriers, which imply dissimilar tunneling probabilities of free carriers. 
Due to this in combination with the $n$-type doping of the WSe$_2$ ML at zero bias and the voltage dependence of the PL spectra, we tentatively proposed the following three scenario of carriers tunneling in our device:
(i) the onsets at 0.8~V/-1~V can be attributed to holes and electrons tunneling into the valence and conduction bands of WSe$_2$ ML, respectively.
(ii) the onsets apparent at around $\pm$3~V are related to consecutive tunneling of electrons/holes, which may form electron-hole pairs ($i.e.$ excitons).
Therefore, the EL signal can be measured, which agrees with the presented its integrated intensity in Figure~\ref{fig4}.
(iii) the 4.5~V onset accompanied with the change of the EL spectra (see Figure~\ref{fig2}(c)) probably is due to the formation of many-body complexes.
At high doping levels of holes, the formation of new species is expected, which can be introduced as a collective response of an electron and a sea of free holes. 
Note that an analogous step is expected for the negative bias voltages.
We believe that its absence is affected by too small applied voltages in our case, as has already been reported in the literature~\cite{Withers2015nano}.
Another possibility of a change of the EL spectra at around 4.5 V can be associated with thermal conditions.
High electrical bias implies significant heating of the device, similar to graphene \cite{Freitag2009, Berciaud2010}, resulting in a thermal population of the emitting state that leads to a broadening of the spectra \cite{Sundaram2013}.

Lastly, we want to address the voltage threshold for the EL signal.
The voltage threshold for EL depends on the exciton binding energy and thermal properties of the material, as well as the characteristics of the contact material. 
In principle,  we should observe the EL signal when the applied voltage is equal to the bandgap energy.
However, in the case of TMD MLs, it was demonstrated the EL emission can be apparent even with the bias energy equaled to the energy of the neutral exciton ($E_{\textrm{X}^0}$$\sim$1.733~eV)~\cite{Binder2017, Withers2015nano, Withers2015nature} due to the possible pinning of the Fermi energy at the WSe$_2$ impurity/acceptor level~\cite{Binder2017}.
The found two times larger bias voltages required in our device can be due to the imperfection of the electrical contacts and the thicknesses of the hBN barriers~\cite{Novotny2023, Binder2017}.
In particular, the measured thickness of the bottom hBN layer of 7 layers can be a dominant reason for high tunneling voltages \cite{Grzeszczyk2024, Britnell2012}. 
	
This proves that the excitation mechanisms of the PL and EL techniques are completely different.
In particular, a much larger number of electron-hole pairs can be formed in the EL experiments, which leads to the broader emission spectra as a result of the formed many-body complexes.

\section{Conclusions}
	
Summarising, the optical emission of the tunneling structure based on a \WSe ML is reported. 
The ML is characterised by high quality as evidenced by the presence of many narrow excitonic lines. 
Due to the charge neutrality point being acquired by applying the positive voltage at about 1~V, we found that the examined ML is naturally $n$-type doped. 
The emission lines apparent in the low-temperature PL spectra under applied bias voltage have been identified.
We studied EL in high career concentration regimes, where spectra are dominated by charged excitonic complexes. 
High electrical bias implies the formation of many-body complexes, which can be also accompanied by significant heating of the device due to the tunneling current.
The obtained results confirm that the excitation mechanisms in PL and EL are different because of the various excitonic complexes seen in both types of experiments. 
A deeper understanding of the processes that occur under conditions of high carriers concentration is needed.

\printcredits

\section*{Declaration of competing interest}
The authors declare that they have no known competing financial interests or personal relationships that could have appeared to influence the work reported in this paper.

\section*{Data availability}
The raw data required to reproduce the above findings are available to download from \href{https://doi.org/10.58132/UIIZRV}{https://doi.org/10.58132/UIIZRV}.
	
\section*{Acknowledgments}
We are grateful to Johannes Binder for fruitful discussions.
The work was supported by the National Science Centre, Poland (Grant No. 2022/46/E/ST3/00166), the Ministry of Education (Singapore) through the Research Centre of Excellence program (Grant EDUN C-33-18-279-V12, I-FIM). 
This research was supported by the Ministry of Education, Singapore, under its Academic Research Fund Tier 2 (T2EP50122-0012). 
This material was based upon work supported by the Air Force European Office of Aerospace Research and Development Office of Scientific Research and the Office of Naval Research Global under award number FA8655-21-1-7026. 
K.W. and T.T. acknowledge support from the JSPS KAKENHI (Grant Numbers 21H05233 and 23H02052) and World Premier International Research Center Initiative (WPI), MEXT, Japan.

\bibliographystyle{cas-model2-names}
\bibliography{biblio}

\end{document}